\documentclass[%
reprint,%
]{revtex4-1}

\usepackage{graphicx}
\usepackage{dcolumn}
\usepackage{bm}
\usepackage{epstopdf}
\usepackage{pdfpages}

\begin{document}

\title{Light matter interaction in WS$_{2}$ nanotube-graphene hybrid devices}
\author{John P. Mathew$^{\dag}$}
\author{Gobinath Jegannathan$^{\dag}$}
\author{Sameer Grover}
\author{Pratiksha D. Dongare}
\author{Rudheer D. Bapat}
\author{Bhagyashree A. Chalke}
\author{S. C. Purandare}
\author{Mandar M. Deshmukh}
\email{deshmukh@tifr.res.in}
\affiliation{Department of Condensed Matter Physics and Materials Science, Tata Institute of Fundamental Research, Homi Bhabha Road, Mumbai 400005, India}
\homepage{contributed equally}


\begin{abstract}
We study the light matter interaction in WS$_{2}$ nanotube-graphene hybrid devices. Using scanning photocurrent microscopy we find that by engineering graphene electrodes for WS$_{2}$ nanotubes we can improve the collection of photogenerated carriers. We observe inhomogeneous spatial photocurrent response with an external quantum efficiency of $\sim$1\% at 0 V bias. We show that defects play an important role and can be utilized to enhance and tune photocarrier generation.
\end{abstract}
\maketitle

Heterostructure devices of transition metal dichalcogenides (TMDCs) and graphene have generated considerable research interest recently because of their superior optical and electronic properties.\cite{britnell_strong_2013,withers2014heterostructures}
The semiconducting nature of TMDCs combined with the presence of van Hove singularities in their electronic density of states allows for efficient photon absorption and carrier generation under optical excitation.\cite{carvalho_band_2013} Combining this feature with the high mobility of graphene has led to optoelectronic studies of heterostructure devices comprising graphene and single layer TMDCs.\cite{britnell_strong_2013,carvalho_band_2013,britnell_field-effect_2012,roy_graphene-mos2_2013,ganatra_few-layer_2014,sun_graphene_2014} These devices have exhibited good quantum efficiency for photocurrent generation in the visible range. However, the fabrication of such heterostructures requires multiple exfoliation and transfer steps. TMDC nanotubes\cite{tenne_polyhedral_1992} represent another alternative for such applications; nanowires offer an additional advantage because they can enhance the absorption of light through the formation of optical cavities\cite{kim_design_2014,frey1998optical} and quasi 1D structures are known to enhance light matter interaction by virtue of an enhanced  joint density of states (JDOS).\cite{dresselhaus_single_2002} Silicon and carbon nanotubes have been shown to be promising materials for solar-cell applications.\cite{tian2007coaxial,wei2007double} Similarly, TMDC nanotubes could also allow for large scale integration of on-chip optoelectronic elements. In addition, the curvature of the nanotubes can be used to engineer spin and valley based optoelectronic control in dichalcogenide systems.\cite{xu_spin_2014}  Here, we investigate the photoresponse of WS$_2$ nanotubes with field-effect transistor geometry and the enhanced photoresponse properties of hybrid devices of WS$_{2}$ nanotubes with graphene electrodes. One of the motivations for using graphene electrodes for the nanotube is to modulate the density of carriers in the electrodes and modify the Schottky barrier;\cite{shih2014tuning} the other motivation is to observe the spatial homogeneity of the photoresponse. We investigate the efficiency of these devices for photoconversion and  attempt to understand the role of defects in modifying optoelectronic properties.\cite{allen2009nonuniform}
\begin{figure}
\includegraphics[width=\columnwidth]{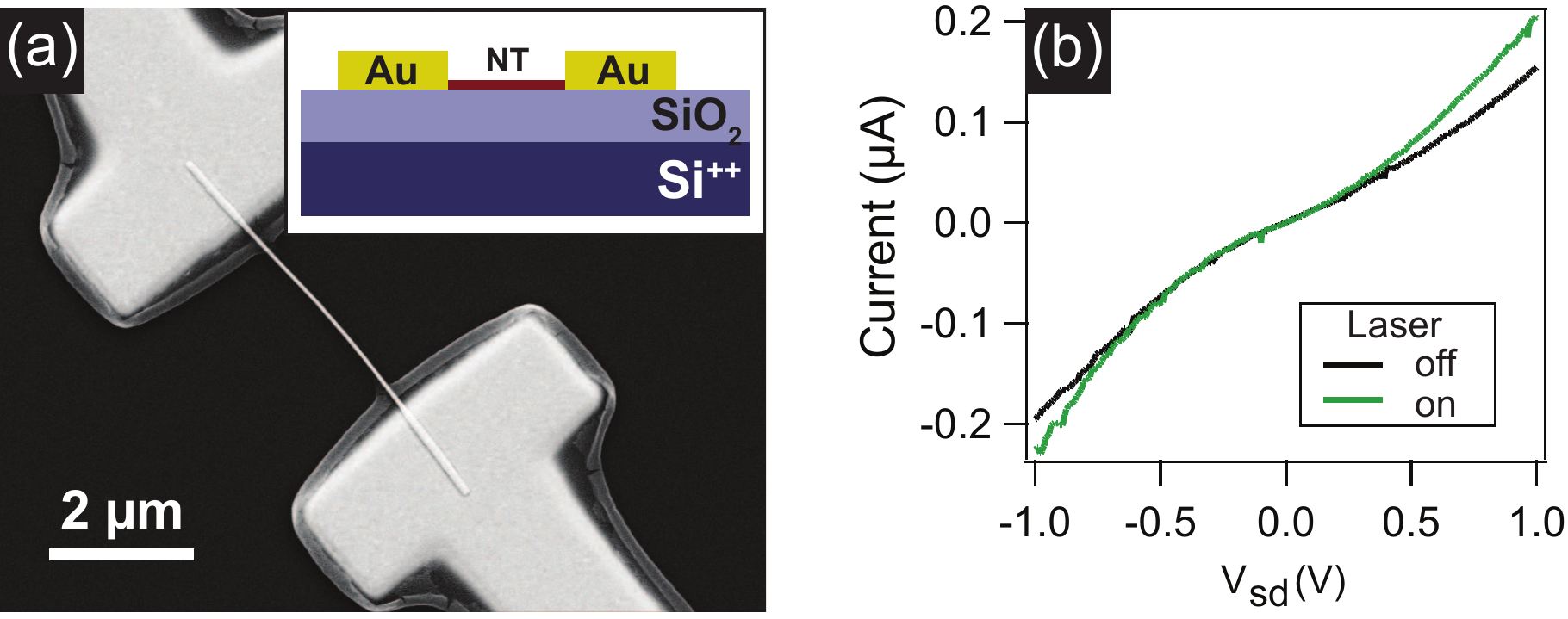}
\caption{ \label{fig:Figure 1} (a) Scanning electron microscope (SEM) image of a WS$_2$ nanotube (NT) device contacted with gold electrodes. (Inset) Schematic diagram of the device in field effect transistor geometry.
(b) I-V characteristics of a WS$_2$ nanotube device measured in ambient conditions at 300 K while applying continuous illumination from a 532 nm laser over the entire device area (laser power $=$ 10 $\mu$W, gate voltage $=$ 0 V). }
\end{figure}

Prior to studying the hybrid devices, we probe individual WS$_{2}$ nanotubes and show that they offer a good optoelectronic platform.\cite{yang2008phototransistors,zhang2012high} We used WS$_2$ multiwalled nanotubes\cite{tenne_polyhedral_1992,levi_field-effect_2013} obtained from NanoMaterials.\cite{apnano} Transmission electron microscope (TEM) images of the nanotubes (see Fig. S1 in supplementary material \cite{suppinfo}) reveal that these are multi-walled hollow tubes of 50$-$200 nm in diameter and have capped, or uncapped ends, where approximately half the diameter is hollow.\cite{tenne_polyhedral_1992} The nanotubes were drop coated on a 300~nm SiO$_2$/Si chip for device fabrication. Standard electron-beam lithography techniques were used for device fabrication. Two probe devices were fabricated by sputtering Au to form metal contacts on the nanotubes as illustrated in Fig. \ref{fig:Figure 1}(a). An in-situ Ar plasma cleaning system was used to remove resist residues under the electrodes before sputtering the metal without breaking the vacuum.

Fig. \ref{fig:Figure 1}(b) shows the I-V characteristics of a WS$_2$ nanotube device measured under ambient conditions. We find that the nanotube forms a Schottky-barrier contact, yielding non-linear current voltage characteristics because of band offsets of WS$_{2}$ and gold.\cite{kang2013band} Although it has been reported\cite{levi_field-effect_2013} that similar devices exhibit n-type behavior, these nanotubes did not exhibit characteristic n or p-type behavior for a back gate voltage ($V_g$)  of up to $\pm$50 V; this may be attributable to a different extent of doping compared with previous studies.\cite{levi_field-effect_2013} In the presence of laser illumination, the device current increased because of photogenerated carriers in the nanotube. The measured photocurrent was observed to be as high as 30 nA in a device with an applied bias of 1 V.

\begin{figure}
\includegraphics[width=\columnwidth]{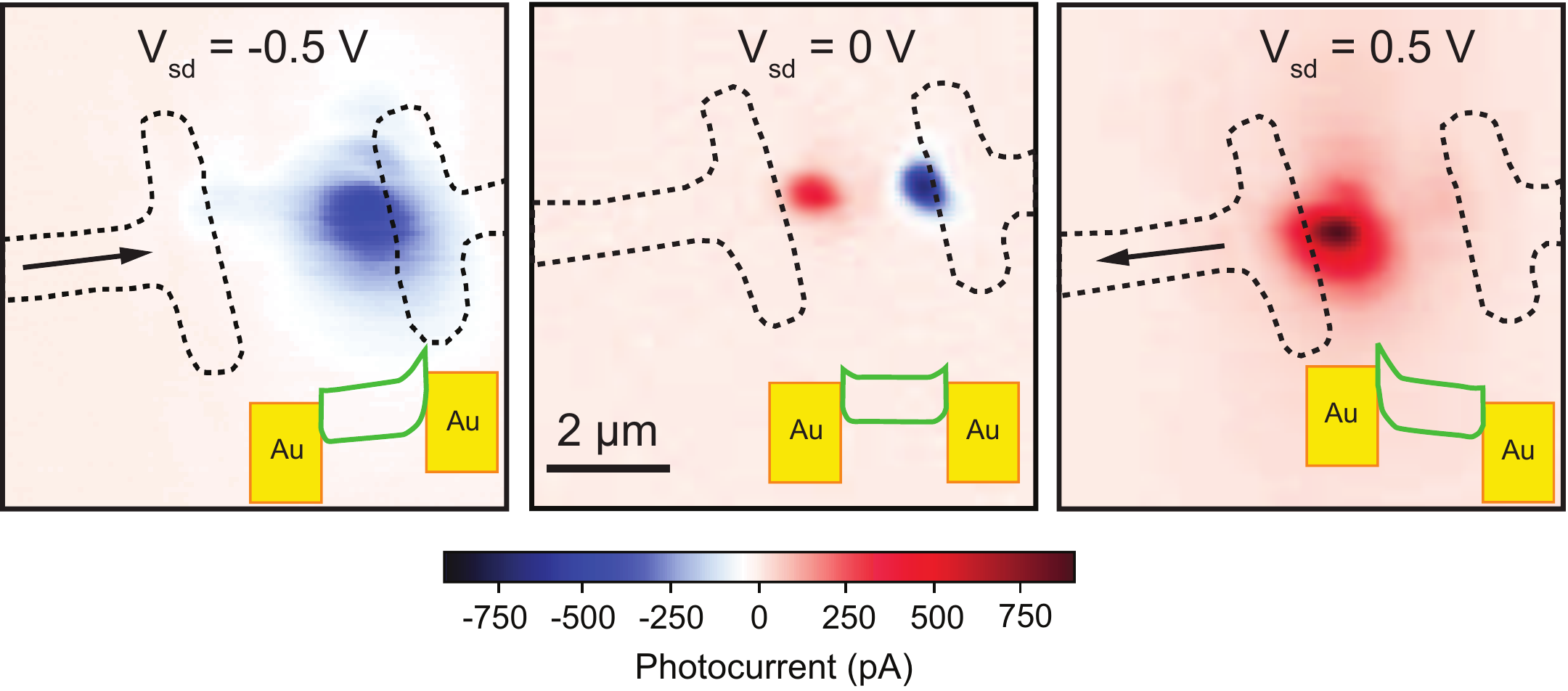}
\caption{\label{fig:Figure 2} Photocurrent response of the WS$_2$ nanotube as a function of the source-drain bias voltage with a gate voltage of 0~V, a 532~nm laser and 1~$\mu$W of power in ambient conditions at 300~K. The applied bias shifts the band bending resulting in a reversal of the photocurrent. The inset in each image illustrates tilting of the bands caused  by the applied bias voltage. The arrows denote the current direction in the device.}
\end{figure}

To further understand the local nature of the photocurrent generation in our devices, we used scanning photocurrent microscopy\cite{lang1978scanning,ahn2005scanning,balasubramanian2004photoelectronic,gu2006quantitative} (see Fig. S3(a) in supplementary material \cite{suppinfo} for a schematic of the measurement setup). We applied a low frequency ($\sim3$ kHz)
modulation to the laser using an acousto-optic modulator (AOM) and a lock-in technique to detect the photocurrent of the devices.  From the response of the nanotube photocurrent to the frequency of modulation of the incident laser (see Fig. S4 in supplementary material \cite{suppinfo}) we deduced a response time of $\sim0.1$ ms for the WS$_{2}$ nanotubes. This result is an improvement compared with the response times of other few/single layer devices of MoS$_{2}$ and WS$_{2}$ reported in the literature\cite{yin2011single,perea2013photosensor,choi2012high} and is comparable to the projected value for WS$_{2}$ nanotubes.\cite{zhang2012high}

The reflected light, incident on the photodetector, was also measured using a lock-in amplifier, which provided information regarding the absorption of light by the material.\cite{zhao_evolution_2013,jiang_electronic_2012} The map of the reflected light (see Fig. \ref{fig:Figure 4}) indicates an absorption of $\sim$ 55$\%$ by the TMDC nanotube; similar values were obtained for WS$_2$ thin films\cite{ballif1996preparation} with an optical absorption coefficient $\alpha=10^{5}$~cm$^{-1}$ (see the supplementary material \cite{suppinfo}).

Fig. \ref{fig:Figure 2}  presents the photocurrent map of a WS$_2$ nanotube device. The electrode positions are outlined using an overlaid map of the reflected signal. We observe that the photocurrent is generated only near the nanotube-metal contact region where the Schottky barrier is present.\cite{sun_graphene_2014,fu_electrothermal_2011,kim2010diameter,ahn2005scanning}  The photocurrent decays into the nanotube away from the Schottky-barrier regions because of carrier diffusion in the nanotube.  An exponential decay function was fitted to the photocurrent profile to extract information about the minority carrier diffusion length, $L_{D}$, in the nanotube.\cite{fu_electrothermal_2011,ubrig2014scanning} The diffusion length was found to be 316 nm from the fit (see Fig. S4(a) in the supplementary material \cite{suppinfo}). The diffusion length is related to the minority charge carrier lifetime, $\tau$, and mobility, $\mu$, by the relation $L_{D}=\sqrt{\frac{k_{B} T}{e}\mu\tau}$, where $k_{B}$ is the Boltzmann constant, $T$ is the temperature and $e$ is the electronic charge. Using previously reported value of hole mobility \cite{braga2012quantitative} in WS$_{2}$, we find the carrier lifetime to be $\tau=0.6$ ns at room temperature. This result is comparable to the carrier lifetimes reported in the literature for other TMDC systems.\cite{shi2013exciton,mouri2013tunable,choi2012high}

The photoresponse as a function of the bias exhibits a shift in the photocurrent toward one of the electrodes with a change in the sign of the bias voltage. The applied bias voltage changes the Schottky-barrier heights of the electrodes asymmetrically, evoking a photoresponse from either end of the nanotube.\cite{freitag2007imaging,lee2007electronic} Photocurrent maps were also generated as a function of the back gate voltage. However, similar to the electrical response, the photoresponse was also observed to be relatively insensitive to the gate voltage.
\begin{figure}
\includegraphics[width=\columnwidth]{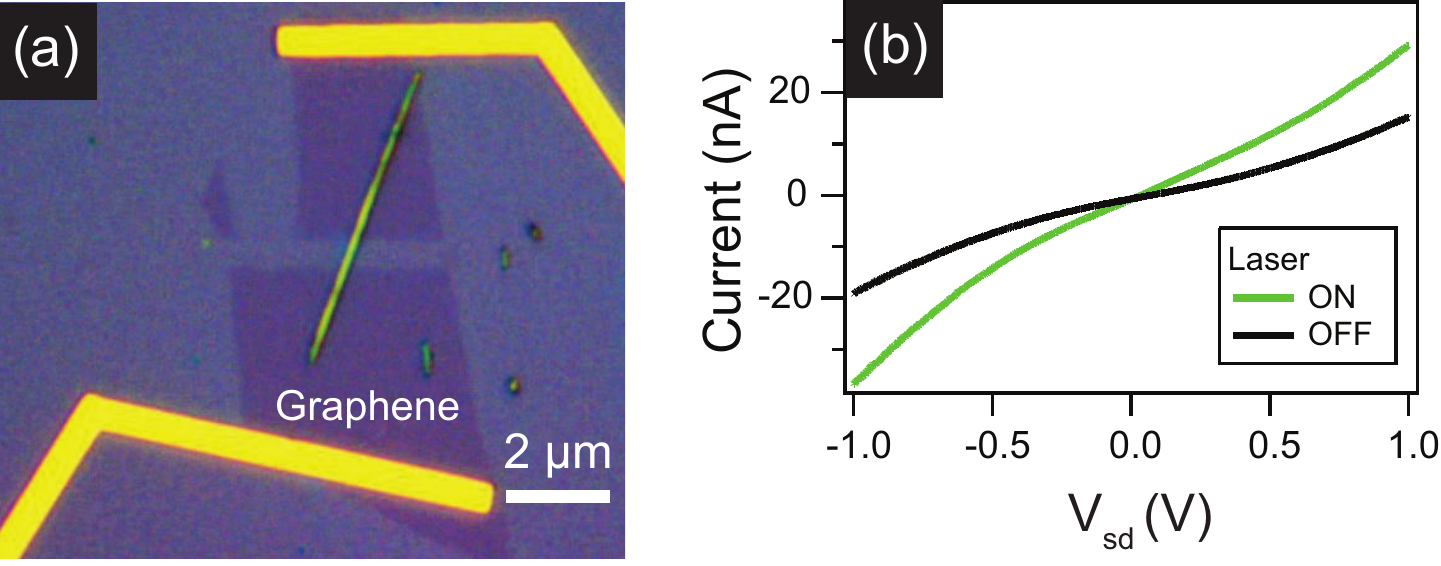}
\caption{\label{fig:Figure 3} (a) Optical image of a WS$_2$ nanotube contacted by graphene electrodes. The slit gap is 1 $\mu$m. (b) I-V characteristics of the device depicted in (a) measured under ambient conditions at 300~K using continuous illumination from a  532 nm laser over the entire device area (laser power $=$ 10 $\mu$W, gate voltage $=$ 0~V.)}
\end{figure}

\begin{figure*}
\center
\includegraphics[width=120mm]{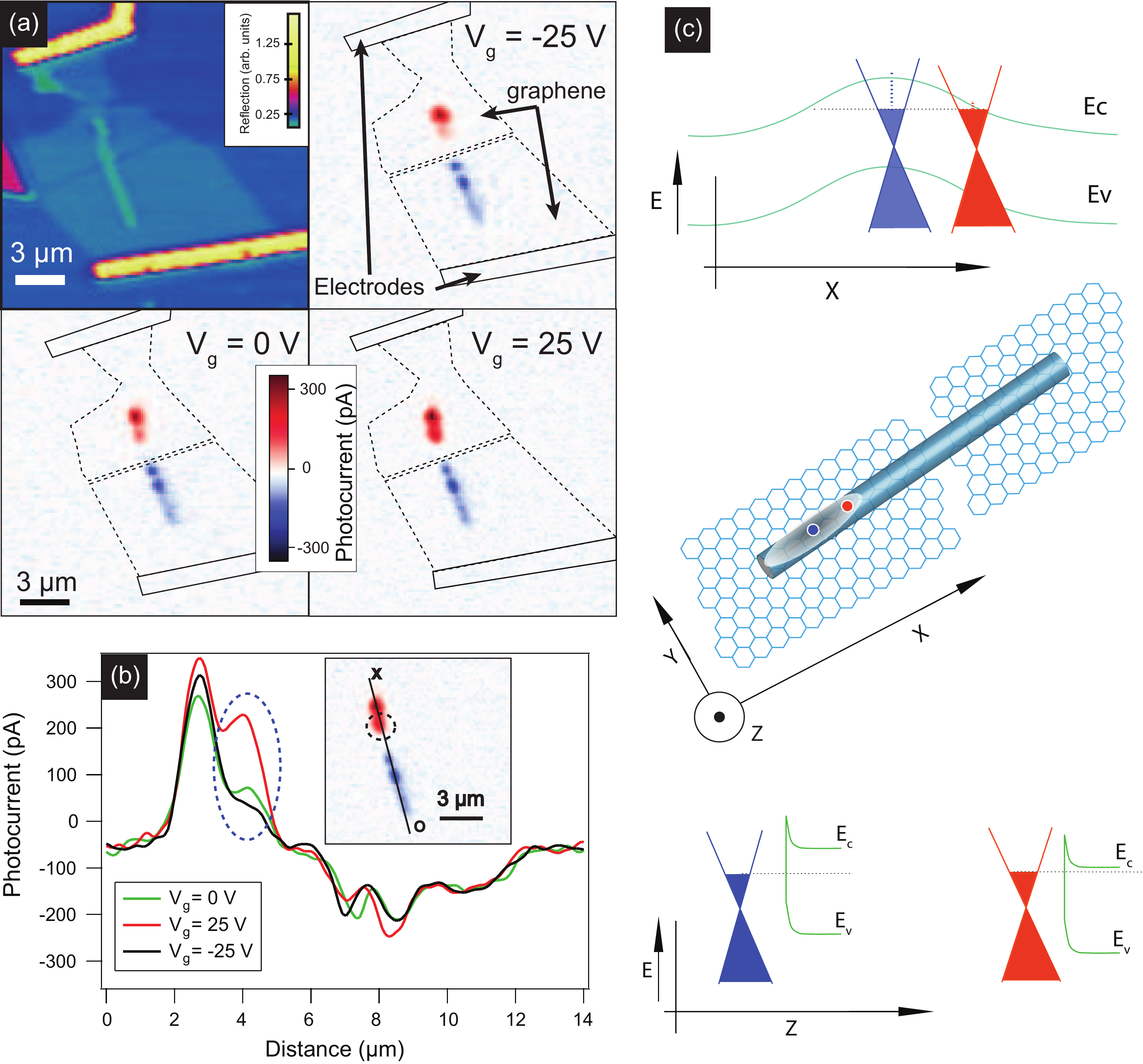}
\caption{\label{fig:Figure 4} (a) Reflection (top left) and photocurrent maps of another slit device (slit gap = 500 nm) with varying back gate voltages obtained using 532 nm laser under ambient conditions at 300 K (bias voltage $=$ 0 V,   laser power $=$ 1 $\mu$W). The photoresponse is observed only in the region where the nanotube overlaps with the graphene. (b) Variation in the photocurrent determined from the  map shown in the inset across the line joining points X and O. The dotted circles indicate the defect region, which is tuned by the gate voltage. (c) Illustration of the local band bending caused by the presence of a defect region on the nanotube. A defect on the nanotube causes band bending along the length of the nanotube. Two points (red and blue circles) along a single defect region are drawn to illustrate the nonuniform interfacial electric field between the graphene and nanotube. The photogenerated charge carriers are collected by the graphene electrode, thus providing a gate tunable photocurrent. (Data from another device are available in the supplementary material.\cite{suppinfo})}
\end{figure*}

Combining the strong light matter interaction of the nanotubes with the superior electrical properties of graphene offers significant advantages in extraction of the generated photocarriers. Now we discuss the second device geometry involving WS$_{2}$ nanotubes with graphene electrodes to study the effect on field effect and photocarrier collection. The fabrication of these devices involves electron beam lithography to fabricate a two probe device out of exfoliated graphene, followed by selective oxygen plasma etching to remove the middle portion of the graphene device leaving behind a slit. A dry transfer process is used to place a WS$_2$ nanotube across the slit in the graphene (see Fig. S5 in supplementary material \cite{suppinfo}). Experiments were conducted using different slit widths. An optical image of such a device is presented in Fig. \ref{fig:Figure 3}(a). The I-V characteristic of the device presented in Fig. \ref{fig:Figure 3}(b) demonstrate that the electrical behavior is similar to that of a two probe WS$_2$ nanotube device\cite{zhang2012electrical} with metal contacts. We also fabricated similar hybrid devices using exfoliated MoS$_2$ and WS$_2$ nanotubes to explore photocarrier diffusion\cite{kang2013band} at the TMDC junction. However, unlike the devices with WS$_2$ nanotubes placed on graphene, the WS$_2$ nanotube with MoS$_2$ electrodes exhibited poor electrical contact.

Although the field effect on the I-V characteristics remains weak, the quantum efficiency of the photoresponse of the graphene-WS$_2$  hybrid devices is $\sim1\%$ at zero bias voltage (see the supplementary material \cite{suppinfo} for details on quantum efficiency calculation).  Fig. \ref{fig:Figure 4} shows the photoresponse of the graphene contacted WS$_2$ nanotube device as a function of the gate voltage at 0~V bias voltage. We observe an inhomogeneous photoresponse along the length of the nanotube at zero bias voltage. At 0~V $V_g$ we see that photocurrent is generated near the interface between the graphene slit and the nanotube\cite{withers2014heterostructures} as well as at the nanotube ends. This finding hints at the presence of band bending near the nanotube ends possibly due to the capped end (see Fig. S1 in supplementary material\cite{suppinfo} for the TEM images).

As defects in the nanotube also modify the local electronic structure of the nanotube, we expect photocurrent generation where there is additional band bending. This phenomenon is observed in Fig. \ref{fig:Figure 4} as the appearance of an additional photocurrent spot with increasing gate voltage.\cite{freitag2007scanning} These spots exhibit higher photoresponse than do other regions along the length of the nanotube. Several  experiments have probed the consequences of such defects on the spatial variation of  photoresponse of  nanostructures.\cite{fu_electrothermal_2011, allen2009nonuniform,lee2007electronic,freitag2007scanning, freitag2007imaging,balasubramanian2005photocurrent} In these experiments, the  effect of the defect is to create a local hill, or a valley, in the potential landscape.  The photocarriers are generated in the same material that carries the charge carriers to the electrodes. This has the consequence that on either  side of the local potential maxima and minima, the sign of the photocurrent changes as the local electric field changes.

In our experiments, the photocarriers generated in the WS$_{2}$ nanotubes were extracted by the graphene electrodes. The interfacial electric field between a defect in the WS$_{2}$ nanotube and the graphene results in the transfer of charge carriers. Because of the variation of the interfacial electric field across the span of the defect, the photocurrent also exhibits a local peak. There was no change in the sign of the photocurrent in our experiments because the sign of the interfacial electric field between the WS$_{2}$ and the graphene did not change; however, its magnitude did change, reflecting a change in the photocurrent. Similar spots of large photoresponse were also seen in the work of Britnell \emph{et al}.\cite{britnell_strong_2013} Although the microscopic nature of the inhomogeneous photocurrent was not discussed in this reference, we believe that the origin was similar in an analogous device structure.

One possible mechanism that could modify the photoresponse is the modification of the workfunction of graphene,\cite{yu_tuning_2009} as this would result in a change in the interfacial electric field. However, not all spots in the spatial photoresponse map are modified as a function of gate voltage. The absence of a uniform modulation of the photoresponse suggests a more localized source away from the graphene.

\begin{figure}
\includegraphics[width=\columnwidth]{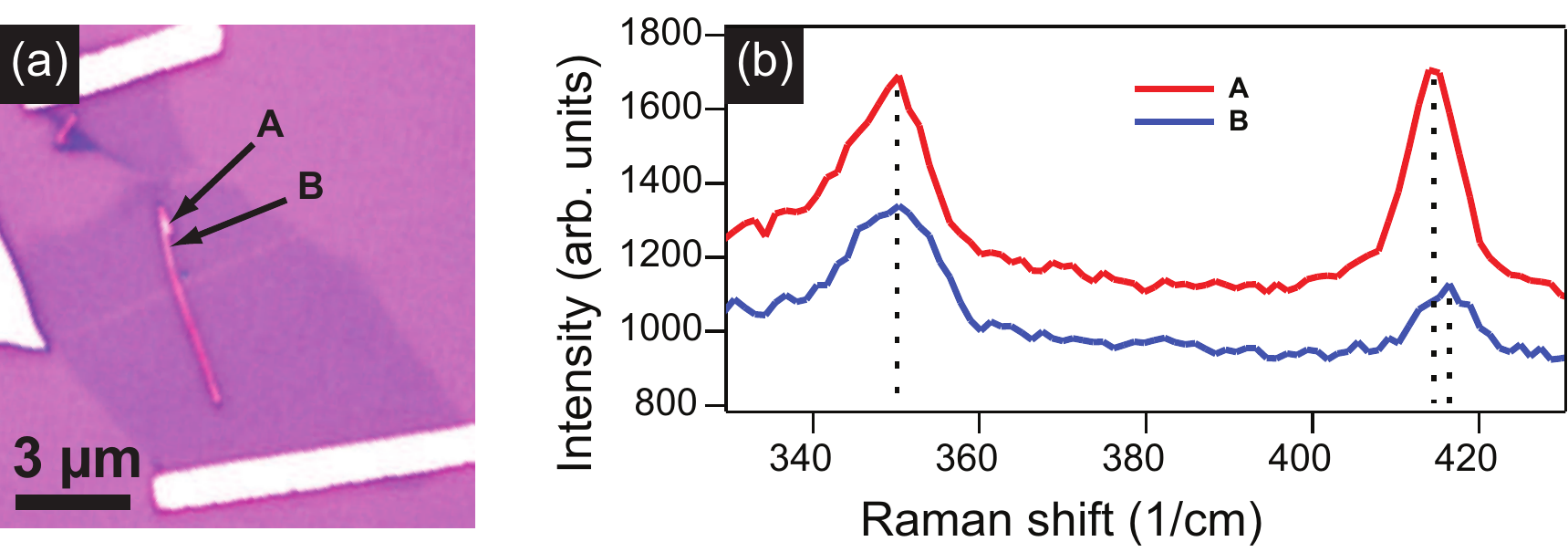}
\caption{\label{fig:Figure 5} (a) Optical image of the device presented in Fig. \ref{fig:Figure 4}. The locations of normal photocurrent (A) and a defect site (B) are marked in the image. (b) Raman spectrum at the site marked A (red curve) and the site marked B (blue curve) in the image on the left for laser excitation at 532 nm. The spectrum at site B exhibits a shift in the characteristic Raman peak of the WS$_2$ nanotube, possibly because of the formation of a crack acting as a defect site (more data are provided in the supplementary material.\cite{suppinfo})}
\end{figure}

A recent study\cite{nan2014strong} reported enhanced photoluminescence in cracked regions of MoS$_2$ monolayers caused by the  adsorption of oxygen in the sulfur vacancy regions accompanied by blue-shift of the $A_{1g}$ mode in the Raman spectrum. Similarly, we observed a blue-shift of $\sim2$~cm$^{-1}$ in the $A_{1g}$ peak of the nanotube at the defect site of the device  (Fig. \ref{fig:Figure 5}).  The formation of these defect regions could be attributable to the ultrasonication of the nanotube solution or to the transfer process and also supports the role of the defects in the enhanced photoresponse observed in our graphene-WS$_2$ devices. However, the resolution of our technique remains limited by the spot size of the incident laser and does not provide any direct insight regarding the microscopic nature of the defects. The presence of localized defects in carbon nanotubes\cite{freitag2007scanning,balasubramanian2005photocurrent, bockrath_resonant_2001} has been studied extensively in the past, and similar defects are likely to exist in inorganic TMDC nanotubes. Our measurement provides a method of imaging defects in similar TMDC semiconductor systems.

In summary we fabricated hybrid devices using WS$_{2}$ nanotubes and graphene that show good light matter interaction. We observed spatially inhomogeneous photoresponse in the nanotube and demonstrated a method of detecting defects in WS$_2$ nanotubes using graphene contacts and scanning photocurrent microscopy. In addition to detecting defects we also observed a good external quantum efficiency ($\sim1\%$) and enhanced photocarrier collection at zero bias voltage in these hybrid devices. The fabrication is simple and does not involve any post-processing because of the robustness of both graphene and WS$_2$ nanotubes. The technique also allows for integration, as WS$_{2}$ nanotubes can be synthesized in large quantities. The strong light matter interaction in dichalcogenide systems and ability of nanowire based structures to concentrate light offer opportunities for fabricating improved photoactive devices using dichalcogenide nanotubes.

We acknowledge fruitful discussion with Aniruddha Konar, IBM. This work was supported by the Swarnajayanti Fellowship from the Department of Science and Technology and, the Department of Atomic Energy of the Government of India.

\pagebreak


\begin{thebibliography}{46}%
\makeatletter
\providecommand \@ifxundefined [1]{%
 \@ifx{#1\undefined}
}%
\providecommand \@ifnum [1]{%
 \ifnum #1\expandafter \@firstoftwo
 \else \expandafter \@secondoftwo
 \fi
}%
\providecommand \@ifx [1]{%
 \ifx #1\expandafter \@firstoftwo
 \else \expandafter \@secondoftwo
 \fi
}%
\providecommand \natexlab [1]{#1}%
\providecommand \enquote  [1]{``#1''}%
\providecommand \bibnamefont  [1]{#1}%
\providecommand \bibfnamefont [1]{#1}%
\providecommand \citenamefont [1]{#1}%
\providecommand \href@noop [0]{\@secondoftwo}%
\providecommand \href [0]{\begingroup \@sanitize@url \@href}%
\providecommand \@href[1]{\@@startlink{#1}\@@href}%
\providecommand \@@href[1]{\endgroup#1\@@endlink}%
\providecommand \@sanitize@url [0]{\catcode `\\12\catcode `\$12\catcode
  `\&12\catcode `\#12\catcode `\^12\catcode `\_12\catcode `\%12\relax}%
\providecommand \@@startlink[1]{}%
\providecommand \@@endlink[0]{}%
\providecommand \url  [0]{\begingroup\@sanitize@url \@url }%
\providecommand \@url [1]{\endgroup\@href {#1}{\urlprefix }}%
\providecommand \urlprefix  [0]{URL }%
\providecommand \Eprint [0]{\href }%
\providecommand \doibase [0]{http://dx.doi.org/}%
\providecommand \selectlanguage [0]{\@gobble}%
\providecommand \bibinfo  [0]{\@secondoftwo}%
\providecommand \bibfield  [0]{\@secondoftwo}%
\providecommand \translation [1]{[#1]}%
\providecommand \BibitemOpen [0]{}%
\providecommand \bibitemStop [0]{}%
\providecommand \bibitemNoStop [0]{.\EOS\space}%
\providecommand \EOS [0]{\spacefactor3000\relax}%
\providecommand \BibitemShut  [1]{\csname bibitem#1\endcsname}%
\let\auto@bib@innerbib\@empty
\bibitem [{\citenamefont {Britnell}\ \emph {et~al.}(2013)\citenamefont
  {Britnell}, \citenamefont {Ribeiro}, \citenamefont {Eckmann}, \citenamefont
  {Jalil}, \citenamefont {Belle}, \citenamefont {Mishchenko}, \citenamefont
  {Kim}, \citenamefont {Gorbachev}, \citenamefont {Georgiou}, \citenamefont
  {Morozov}, \citenamefont {Grigorenko}, \citenamefont {Geim}, \citenamefont
  {Casiraghi}, \citenamefont {Castro~Neto},\ and\ \citenamefont
  {Novoselov}}]{britnell_strong_2013}%
  \BibitemOpen
  \bibfield  {author} {\bibinfo {author} {\bibfnamefont {L.}~\bibnamefont
  {Britnell}}, \bibinfo {author} {\bibfnamefont {R.}~\bibnamefont {Ribeiro}},
  \bibinfo {author} {\bibfnamefont {A.}~\bibnamefont {Eckmann}}, \bibinfo
  {author} {\bibfnamefont {R.}~\bibnamefont {Jalil}}, \bibinfo {author}
  {\bibfnamefont {B.}~\bibnamefont {Belle}}, \bibinfo {author} {\bibfnamefont
  {A.}~\bibnamefont {Mishchenko}}, \bibinfo {author} {\bibfnamefont {Y.-J.}\
  \bibnamefont {Kim}}, \bibinfo {author} {\bibfnamefont {R.}~\bibnamefont
  {Gorbachev}}, \bibinfo {author} {\bibfnamefont {T.}~\bibnamefont {Georgiou}},
  \bibinfo {author} {\bibfnamefont {S.}~\bibnamefont {Morozov}}, \bibinfo
  {author} {\bibfnamefont {A.~N.}\ \bibnamefont {Grigorenko}}, \bibinfo
  {author} {\bibfnamefont {A.~K.}\ \bibnamefont {Geim}}, \bibinfo {author}
  {\bibfnamefont {C.}~\bibnamefont {Casiraghi}}, \bibinfo {author}
  {\bibfnamefont {A.~H.}\ \bibnamefont {Castro~Neto}}, \ and\ \bibinfo {author}
  {\bibfnamefont {K.~S.}\ \bibnamefont {Novoselov}},\ }\href@noop {} {\bibfield
   {journal} {\bibinfo  {journal} {Science}\ }\textbf {\bibinfo {volume}
  {340}},\ \bibinfo {pages} {1311} (\bibinfo {year} {2013})}\BibitemShut
  {NoStop}%
\bibitem [{\citenamefont {Withers}\ \emph {et~al.}(2014)\citenamefont
  {Withers}, \citenamefont {Yang}, \citenamefont {Britnell}, \citenamefont
  {Rooney}, \citenamefont {Lewis}, \citenamefont {Felten}, \citenamefont
  {Woods}, \citenamefont {Sanchez~Romaguera}, \citenamefont {Georgiou},
  \citenamefont {Eckmann}, \citenamefont {Kim}, \citenamefont {Yeates},
  \citenamefont {Haigh}, \citenamefont {Geim}, \citenamefont {Novoselov},\ and\
  \citenamefont {Casiraghi}}]{withers2014heterostructures}%
  \BibitemOpen
  \bibfield  {author} {\bibinfo {author} {\bibfnamefont {F.}~\bibnamefont
  {Withers}}, \bibinfo {author} {\bibfnamefont {H.}~\bibnamefont {Yang}},
  \bibinfo {author} {\bibfnamefont {L.}~\bibnamefont {Britnell}}, \bibinfo
  {author} {\bibfnamefont {A.~P.}\ \bibnamefont {Rooney}}, \bibinfo {author}
  {\bibfnamefont {E.}~\bibnamefont {Lewis}}, \bibinfo {author} {\bibfnamefont
  {A.}~\bibnamefont {Felten}}, \bibinfo {author} {\bibfnamefont {C.~R.}\
  \bibnamefont {Woods}}, \bibinfo {author} {\bibfnamefont {V.}~\bibnamefont
  {Sanchez~Romaguera}}, \bibinfo {author} {\bibfnamefont {T.}~\bibnamefont
  {Georgiou}}, \bibinfo {author} {\bibfnamefont {A.}~\bibnamefont {Eckmann}},
  \bibinfo {author} {\bibfnamefont {Y.~J.}\ \bibnamefont {Kim}}, \bibinfo
  {author} {\bibfnamefont {S.~G.}\ \bibnamefont {Yeates}}, \bibinfo {author}
  {\bibfnamefont {S.~J.}\ \bibnamefont {Haigh}}, \bibinfo {author}
  {\bibfnamefont {A.~K.}\ \bibnamefont {Geim}}, \bibinfo {author}
  {\bibfnamefont {K.~S.}\ \bibnamefont {Novoselov}}, \ and\ \bibinfo {author}
  {\bibfnamefont {C.}~\bibnamefont {Casiraghi}},\ }\href {\doibase
  10.1021/nl501355j} {\bibfield  {journal} {\bibinfo  {journal} {Nano Letters}\
  }\textbf {\bibinfo {volume} {14}},\ \bibinfo {pages} {3987} (\bibinfo {year}
  {2014})}\BibitemShut {NoStop}%
\bibitem [{\citenamefont {Carvalho}\ \emph {et~al.}(2013)\citenamefont
  {Carvalho}, \citenamefont {Ribeiro},\ and\ \citenamefont
  {Castro~Neto}}]{carvalho_band_2013}%
  \BibitemOpen
  \bibfield  {author} {\bibinfo {author} {\bibfnamefont {A.}~\bibnamefont
  {Carvalho}}, \bibinfo {author} {\bibfnamefont {R.~M.}\ \bibnamefont
  {Ribeiro}}, \ and\ \bibinfo {author} {\bibfnamefont {A.~H.}\ \bibnamefont
  {Castro~Neto}},\ }\href {\doibase 10.1103/PhysRevB.88.115205} {\bibfield
  {journal} {\bibinfo  {journal} {Physical Review B}\ }\textbf {\bibinfo
  {volume} {88}},\ \bibinfo {pages} {115205} (\bibinfo {year}
  {2013})}\BibitemShut {NoStop}%
\bibitem [{\citenamefont {Britnell}\ \emph {et~al.}(2012)\citenamefont
  {Britnell}, \citenamefont {Gorbachev}, \citenamefont {Jalil}, \citenamefont
  {Belle}, \citenamefont {Schedin}, \citenamefont {Mishchenko}, \citenamefont
  {Georgiou}, \citenamefont {Katsnelson}, \citenamefont {Eaves}, \citenamefont
  {Morozov}, \citenamefont {Peres}, \citenamefont {Leist}, \citenamefont
  {Geim}, \citenamefont {Novoselov},\ and\ \citenamefont
  {Ponomarenko}}]{britnell_field-effect_2012}%
  \BibitemOpen
  \bibfield  {author} {\bibinfo {author} {\bibfnamefont {L.}~\bibnamefont
  {Britnell}}, \bibinfo {author} {\bibfnamefont {R.~V.}\ \bibnamefont
  {Gorbachev}}, \bibinfo {author} {\bibfnamefont {R.}~\bibnamefont {Jalil}},
  \bibinfo {author} {\bibfnamefont {B.~D.}\ \bibnamefont {Belle}}, \bibinfo
  {author} {\bibfnamefont {F.}~\bibnamefont {Schedin}}, \bibinfo {author}
  {\bibfnamefont {A.}~\bibnamefont {Mishchenko}}, \bibinfo {author}
  {\bibfnamefont {T.}~\bibnamefont {Georgiou}}, \bibinfo {author}
  {\bibfnamefont {M.~I.}\ \bibnamefont {Katsnelson}}, \bibinfo {author}
  {\bibfnamefont {L.}~\bibnamefont {Eaves}}, \bibinfo {author} {\bibfnamefont
  {S.~V.}\ \bibnamefont {Morozov}}, \bibinfo {author} {\bibfnamefont
  {N.~M.~R.}\ \bibnamefont {Peres}}, \bibinfo {author} {\bibfnamefont
  {J.}~\bibnamefont {Leist}}, \bibinfo {author} {\bibfnamefont {A.~K.}\
  \bibnamefont {Geim}}, \bibinfo {author} {\bibfnamefont {K.~S.}\ \bibnamefont
  {Novoselov}}, \ and\ \bibinfo {author} {\bibfnamefont {L.~A.}\ \bibnamefont
  {Ponomarenko}},\ }\href {\doibase 10.1126/science.1218461} {\bibfield
  {journal} {\bibinfo  {journal} {Science}\ }\textbf {\bibinfo {volume}
  {335}},\ \bibinfo {pages} {947} (\bibinfo {year} {2012})}\BibitemShut
  {NoStop}%
\bibitem [{\citenamefont {Roy}\ \emph {et~al.}(2013)\citenamefont {Roy},
  \citenamefont {Padmanabhan}, \citenamefont {Goswami}, \citenamefont {Sai},
  \citenamefont {Ramalingam}, \citenamefont {Raghavan},\ and\ \citenamefont
  {Ghosh}}]{roy_graphene-mos2_2013}%
  \BibitemOpen
  \bibfield  {author} {\bibinfo {author} {\bibfnamefont {K.}~\bibnamefont
  {Roy}}, \bibinfo {author} {\bibfnamefont {M.}~\bibnamefont {Padmanabhan}},
  \bibinfo {author} {\bibfnamefont {S.}~\bibnamefont {Goswami}}, \bibinfo
  {author} {\bibfnamefont {T.~P.}\ \bibnamefont {Sai}}, \bibinfo {author}
  {\bibfnamefont {G.}~\bibnamefont {Ramalingam}}, \bibinfo {author}
  {\bibfnamefont {S.}~\bibnamefont {Raghavan}}, \ and\ \bibinfo {author}
  {\bibfnamefont {A.}~\bibnamefont {Ghosh}},\ }\href {\doibase
  10.1038/nnano.2013.206} {\bibfield  {journal} {\bibinfo  {journal} {Nature
  Nanotechnology}\ }\textbf {\bibinfo {volume} {8}},\ \bibinfo {pages} {826}
  (\bibinfo {year} {2013})}\BibitemShut {NoStop}%
\bibitem [{\citenamefont {Ganatra}\ and\ \citenamefont
  {Zhang}(2014)}]{ganatra_few-layer_2014}%
  \BibitemOpen
  \bibfield  {author} {\bibinfo {author} {\bibfnamefont {R.}~\bibnamefont
  {Ganatra}}\ and\ \bibinfo {author} {\bibfnamefont {Q.}~\bibnamefont
  {Zhang}},\ }\href {\doibase 10.1021/nn405938z} {\bibfield  {journal}
  {\bibinfo  {journal} {ACS Nano}\ }\textbf {\bibinfo {volume} {8}},\ \bibinfo
  {pages} {4074} (\bibinfo {year} {2014})}\BibitemShut {NoStop}%
\bibitem [{\citenamefont {Sun}\ and\ \citenamefont
  {Chang}(2014)}]{sun_graphene_2014}%
  \BibitemOpen
  \bibfield  {author} {\bibinfo {author} {\bibfnamefont {Z.}~\bibnamefont
  {Sun}}\ and\ \bibinfo {author} {\bibfnamefont {H.}~\bibnamefont {Chang}},\
  }\href {\doibase 10.1021/nn500508c} {\bibfield  {journal} {\bibinfo
  {journal} {ACS Nano}\ }\textbf {\bibinfo {volume} {8}},\ \bibinfo {pages}
  {4133} (\bibinfo {year} {2014})}\BibitemShut {NoStop}%
\bibitem [{\citenamefont {Tenne}\ \emph {et~al.}(1992)\citenamefont {Tenne},
  \citenamefont {Margulis}, \citenamefont {Genut},\ and\ \citenamefont
  {Hodes}}]{tenne_polyhedral_1992}%
  \BibitemOpen
  \bibfield  {author} {\bibinfo {author} {\bibfnamefont {R.}~\bibnamefont
  {Tenne}}, \bibinfo {author} {\bibfnamefont {L.}~\bibnamefont {Margulis}},
  \bibinfo {author} {\bibfnamefont {M.}~\bibnamefont {Genut}}, \ and\ \bibinfo
  {author} {\bibfnamefont {G.}~\bibnamefont {Hodes}},\ }\href {\doibase
  10.1038/360444a0} {\bibfield  {journal} {\bibinfo  {journal} {Nature}\
  }\textbf {\bibinfo {volume} {360}},\ \bibinfo {pages} {444} (\bibinfo {year}
  {1992})}\BibitemShut {NoStop}%
\bibitem [{\citenamefont {Kim}\ \emph {et~al.}(2014)\citenamefont {Kim},
  \citenamefont {Song}, \citenamefont {Kempa}, \citenamefont {Day},
  \citenamefont {Lieber},\ and\ \citenamefont {Park}}]{kim_design_2014}%
  \BibitemOpen
  \bibfield  {author} {\bibinfo {author} {\bibfnamefont {S.-K.}\ \bibnamefont
  {Kim}}, \bibinfo {author} {\bibfnamefont {K.-D.}\ \bibnamefont {Song}},
  \bibinfo {author} {\bibfnamefont {T.~J.}\ \bibnamefont {Kempa}}, \bibinfo
  {author} {\bibfnamefont {R.~W.}\ \bibnamefont {Day}}, \bibinfo {author}
  {\bibfnamefont {C.~M.}\ \bibnamefont {Lieber}}, \ and\ \bibinfo {author}
  {\bibfnamefont {H.-G.}\ \bibnamefont {Park}},\ }\href {\doibase
  10.1021/nn5003776} {\bibfield  {journal} {\bibinfo  {journal} {{ACS} Nano}\
  }\textbf {\bibinfo {volume} {8}},\ \bibinfo {pages} {3707} (\bibinfo {year}
  {2014})}\BibitemShut {NoStop}%
\bibitem [{\citenamefont {Frey}\ \emph {et~al.}(1998)\citenamefont {Frey},
  \citenamefont {Tenne}, \citenamefont {Matthews}, \citenamefont
  {Dresselhaus},\ and\ \citenamefont {Dresselhaus}}]{frey1998optical}%
  \BibitemOpen
  \bibfield  {author} {\bibinfo {author} {\bibfnamefont {G.}~\bibnamefont
  {Frey}}, \bibinfo {author} {\bibfnamefont {R.}~\bibnamefont {Tenne}},
  \bibinfo {author} {\bibfnamefont {M.}~\bibnamefont {Matthews}}, \bibinfo
  {author} {\bibfnamefont {M.}~\bibnamefont {Dresselhaus}}, \ and\ \bibinfo
  {author} {\bibfnamefont {G.}~\bibnamefont {Dresselhaus}},\ }\href@noop {}
  {\bibfield  {journal} {\bibinfo  {journal} {Journal of Materials Research}\
  }\textbf {\bibinfo {volume} {13}},\ \bibinfo {pages} {2412} (\bibinfo {year}
  {1998})}\BibitemShut {NoStop}%
\bibitem [{\citenamefont {Dresselhaus}\ \emph {et~al.}(2002)\citenamefont
  {Dresselhaus}, \citenamefont {Dresselhaus}, \citenamefont {Jorio},
  \citenamefont {Souza~Filho}, \citenamefont {Pimenta},\ and\ \citenamefont
  {Saito}}]{dresselhaus_single_2002}%
  \BibitemOpen
  \bibfield  {author} {\bibinfo {author} {\bibfnamefont {M.~S.}\ \bibnamefont
  {Dresselhaus}}, \bibinfo {author} {\bibfnamefont {G.}~\bibnamefont
  {Dresselhaus}}, \bibinfo {author} {\bibfnamefont {A.}~\bibnamefont {Jorio}},
  \bibinfo {author} {\bibfnamefont {A.~G.}\ \bibnamefont {Souza~Filho}},
  \bibinfo {author} {\bibfnamefont {M.~A.}\ \bibnamefont {Pimenta}}, \ and\
  \bibinfo {author} {\bibfnamefont {R.}~\bibnamefont {Saito}},\ }\href
  {\doibase 10.1021/ar0101537} {\bibfield  {journal} {\bibinfo  {journal}
  {Accounts of Chemical Research}\ }\textbf {\bibinfo {volume} {35}},\ \bibinfo
  {pages} {1070} (\bibinfo {year} {2002})}\BibitemShut {NoStop}%
\bibitem [{\citenamefont {Tian}\ \emph {et~al.}(2007)\citenamefont {Tian},
  \citenamefont {Zheng}, \citenamefont {Kempa}, \citenamefont {Fang},
  \citenamefont {Yu}, \citenamefont {Yu}, \citenamefont {Huang},\ and\
  \citenamefont {Lieber}}]{tian2007coaxial}%
  \BibitemOpen
  \bibfield  {author} {\bibinfo {author} {\bibfnamefont {B.}~\bibnamefont
  {Tian}}, \bibinfo {author} {\bibfnamefont {X.}~\bibnamefont {Zheng}},
  \bibinfo {author} {\bibfnamefont {T.~J.}\ \bibnamefont {Kempa}}, \bibinfo
  {author} {\bibfnamefont {Y.}~\bibnamefont {Fang}}, \bibinfo {author}
  {\bibfnamefont {N.}~\bibnamefont {Yu}}, \bibinfo {author} {\bibfnamefont
  {G.}~\bibnamefont {Yu}}, \bibinfo {author} {\bibfnamefont {J.}~\bibnamefont
  {Huang}}, \ and\ \bibinfo {author} {\bibfnamefont {C.~M.}\ \bibnamefont
  {Lieber}},\ }\href@noop {} {\bibfield  {journal} {\bibinfo  {journal}
  {Nature}\ }\textbf {\bibinfo {volume} {449}},\ \bibinfo {pages} {885}
  (\bibinfo {year} {2007})}\BibitemShut {NoStop}%
\bibitem [{\citenamefont {Wei}\ \emph {et~al.}(2007)\citenamefont {Wei},
  \citenamefont {Jia}, \citenamefont {Shu}, \citenamefont {Gu}, \citenamefont
  {Wang}, \citenamefont {Zhuang}, \citenamefont {Zhang}, \citenamefont {Wang},
  \citenamefont {Luo}, \citenamefont {Cao},\ and\ \citenamefont
  {Wu}}]{wei2007double}%
  \BibitemOpen
  \bibfield  {author} {\bibinfo {author} {\bibfnamefont {J.}~\bibnamefont
  {Wei}}, \bibinfo {author} {\bibfnamefont {Y.}~\bibnamefont {Jia}}, \bibinfo
  {author} {\bibfnamefont {Q.}~\bibnamefont {Shu}}, \bibinfo {author}
  {\bibfnamefont {Z.}~\bibnamefont {Gu}}, \bibinfo {author} {\bibfnamefont
  {K.}~\bibnamefont {Wang}}, \bibinfo {author} {\bibfnamefont {D.}~\bibnamefont
  {Zhuang}}, \bibinfo {author} {\bibfnamefont {G.}~\bibnamefont {Zhang}},
  \bibinfo {author} {\bibfnamefont {Z.}~\bibnamefont {Wang}}, \bibinfo {author}
  {\bibfnamefont {J.}~\bibnamefont {Luo}}, \bibinfo {author} {\bibfnamefont
  {A.}~\bibnamefont {Cao}}, \ and\ \bibinfo {author} {\bibfnamefont
  {D.}~\bibnamefont {Wu}},\ }\href@noop {} {\bibfield  {journal} {\bibinfo
  {journal} {Nano Letters}\ }\textbf {\bibinfo {volume} {7}},\ \bibinfo {pages}
  {2317} (\bibinfo {year} {2007})}\BibitemShut {NoStop}%
\bibitem [{\citenamefont {Xu}\ \emph {et~al.}(2014)\citenamefont {Xu},
  \citenamefont {Yao}, \citenamefont {Xiao},\ and\ \citenamefont
  {Heinz}}]{xu_spin_2014}%
  \BibitemOpen
  \bibfield  {author} {\bibinfo {author} {\bibfnamefont {X.}~\bibnamefont
  {Xu}}, \bibinfo {author} {\bibfnamefont {W.}~\bibnamefont {Yao}}, \bibinfo
  {author} {\bibfnamefont {D.}~\bibnamefont {Xiao}}, \ and\ \bibinfo {author}
  {\bibfnamefont {T.~F.}\ \bibnamefont {Heinz}},\ }\href {\doibase
  10.1038/nphys2942} {\bibfield  {journal} {\bibinfo  {journal} {Nature
  Physics}\ }\textbf {\bibinfo {volume} {10}},\ \bibinfo {pages} {343}
  (\bibinfo {year} {2014})}\BibitemShut {NoStop}%
\bibitem [{\citenamefont {Shih}\ \emph {et~al.}(2014)\citenamefont {Shih},
  \citenamefont {Wang}, \citenamefont {Son}, \citenamefont {Jin}, \citenamefont
  {Blankschtein},\ and\ \citenamefont {Strano}}]{shih2014tuning}%
  \BibitemOpen
  \bibfield  {author} {\bibinfo {author} {\bibfnamefont {C.-J.}\ \bibnamefont
  {Shih}}, \bibinfo {author} {\bibfnamefont {Q.~H.}\ \bibnamefont {Wang}},
  \bibinfo {author} {\bibfnamefont {Y.}~\bibnamefont {Son}}, \bibinfo {author}
  {\bibfnamefont {Z.}~\bibnamefont {Jin}}, \bibinfo {author} {\bibfnamefont
  {D.}~\bibnamefont {Blankschtein}}, \ and\ \bibinfo {author} {\bibfnamefont
  {M.~S.}\ \bibnamefont {Strano}},\ }\href {\doibase 10.1021/nn500676t}
  {\bibfield  {journal} {\bibinfo  {journal} {ACS Nano}\ }\textbf {\bibinfo
  {volume} {8}},\ \bibinfo {pages} {5790} (\bibinfo {year} {2014})}\BibitemShut
  {NoStop}%
\bibitem [{\citenamefont {Allen}\ \emph {et~al.}(2009)\citenamefont {Allen},
  \citenamefont {Perea}, \citenamefont {Hemesath},\ and\ \citenamefont
  {Lauhon}}]{allen2009nonuniform}%
  \BibitemOpen
  \bibfield  {author} {\bibinfo {author} {\bibfnamefont {J.~E.}\ \bibnamefont
  {Allen}}, \bibinfo {author} {\bibfnamefont {D.~E.}\ \bibnamefont {Perea}},
  \bibinfo {author} {\bibfnamefont {E.~R.}\ \bibnamefont {Hemesath}}, \ and\
  \bibinfo {author} {\bibfnamefont {L.~J.}\ \bibnamefont {Lauhon}},\
  }\href@noop {} {\bibfield  {journal} {\bibinfo  {journal} {Advanced
  Materials}\ }\textbf {\bibinfo {volume} {21}},\ \bibinfo {pages} {3067}
  (\bibinfo {year} {2009})}\BibitemShut {NoStop}%
\bibitem [{\citenamefont {Yang}\ \emph {et~al.}(2008)\citenamefont {Yang},
  \citenamefont {Unalan}, \citenamefont {Hiralal}, \citenamefont {Chremmou},
  \citenamefont {The}, \citenamefont {Alexandrou}, \citenamefont {Tenne},\ and\
  \citenamefont {Amaratunga}}]{yang2008phototransistors}%
  \BibitemOpen
  \bibfield  {author} {\bibinfo {author} {\bibfnamefont {Y.}~\bibnamefont
  {Yang}}, \bibinfo {author} {\bibfnamefont {H.~E.}\ \bibnamefont {Unalan}},
  \bibinfo {author} {\bibfnamefont {P.}~\bibnamefont {Hiralal}}, \bibinfo
  {author} {\bibfnamefont {K.}~\bibnamefont {Chremmou}}, \bibinfo {author}
  {\bibfnamefont {A.}~\bibnamefont {The}}, \bibinfo {author} {\bibfnamefont
  {I.}~\bibnamefont {Alexandrou}}, \bibinfo {author} {\bibfnamefont
  {R.}~\bibnamefont {Tenne}}, \ and\ \bibinfo {author} {\bibfnamefont
  {G.}~\bibnamefont {Amaratunga}},\ }in\ \href@noop {} {\emph {\bibinfo
  {booktitle} {Nanotechnology, 2008. NANO'08. 8th IEEE Conference on}}}\
  (\bibinfo {organization} {IEEE},\ \bibinfo {year} {2008})\ pp.\ \bibinfo
  {pages} {85--87}\BibitemShut {NoStop}%
\bibitem [{\citenamefont {Zhang}\ \emph
  {et~al.}(2012{\natexlab{a}})\citenamefont {Zhang}, \citenamefont {Wang},
  \citenamefont {Yang}, \citenamefont {Liu}, \citenamefont {Xu}, \citenamefont
  {Ning}, \citenamefont {Zak}, \citenamefont {Zhang}, \citenamefont {Tenne},\
  and\ \citenamefont {Chen}}]{zhang2012high}%
  \BibitemOpen
  \bibfield  {author} {\bibinfo {author} {\bibfnamefont {C.}~\bibnamefont
  {Zhang}}, \bibinfo {author} {\bibfnamefont {S.}~\bibnamefont {Wang}},
  \bibinfo {author} {\bibfnamefont {L.}~\bibnamefont {Yang}}, \bibinfo {author}
  {\bibfnamefont {Y.}~\bibnamefont {Liu}}, \bibinfo {author} {\bibfnamefont
  {T.}~\bibnamefont {Xu}}, \bibinfo {author} {\bibfnamefont {Z.}~\bibnamefont
  {Ning}}, \bibinfo {author} {\bibfnamefont {A.}~\bibnamefont {Zak}}, \bibinfo
  {author} {\bibfnamefont {Z.}~\bibnamefont {Zhang}}, \bibinfo {author}
  {\bibfnamefont {R.}~\bibnamefont {Tenne}}, \ and\ \bibinfo {author}
  {\bibfnamefont {Q.}~\bibnamefont {Chen}},\ }\href@noop {} {\bibfield
  {journal} {\bibinfo  {journal} {Applied Physics Letters}\ }\textbf {\bibinfo
  {volume} {100}},\ \bibinfo {pages} {243101} (\bibinfo {year}
  {2012}{\natexlab{a}})}\BibitemShut {NoStop}%
\bibitem [{\citenamefont {Levi}\ \emph {et~al.}(2013)\citenamefont {Levi},
  \citenamefont {Bitton}, \citenamefont {Leitus}, \citenamefont {Tenne},\ and\
  \citenamefont {Joselevich}}]{levi_field-effect_2013}%
  \BibitemOpen
  \bibfield  {author} {\bibinfo {author} {\bibfnamefont {R.}~\bibnamefont
  {Levi}}, \bibinfo {author} {\bibfnamefont {O.}~\bibnamefont {Bitton}},
  \bibinfo {author} {\bibfnamefont {G.}~\bibnamefont {Leitus}}, \bibinfo
  {author} {\bibfnamefont {R.}~\bibnamefont {Tenne}}, \ and\ \bibinfo {author}
  {\bibfnamefont {E.}~\bibnamefont {Joselevich}},\ }\href {\doibase
  10.1021/nl401675k} {\bibfield  {journal} {\bibinfo  {journal} {Nano Letters}\
  }\textbf {\bibinfo {volume} {13}},\ \bibinfo {pages} {3736} (\bibinfo {year}
  {2013})}\BibitemShut {NoStop}%
\bibitem [{apn()}]{apnano}%
  \BibitemOpen
  \href@noop {} {}\bibinfo {note} {NanoMaterials Ltd., Yavne,
  Israel}\BibitemShut {NoStop}%
\bibitem [{sup()}]{suppinfo}%
  \BibitemOpen
  \href@noop {} {}\bibinfo {note} {See supplementary material at http://dx.doi.org/10.1063/1.4902983 for TEM images,
  measurement scheme, and data from another device.}\BibitemShut {Stop}%
\bibitem [{\citenamefont {Kang}\ \emph {et~al.}(2013)\citenamefont {Kang},
  \citenamefont {Tongay}, \citenamefont {Zhou}, \citenamefont {Li},\ and\
  \citenamefont {Wu}}]{kang2013band}%
  \BibitemOpen
  \bibfield  {author} {\bibinfo {author} {\bibfnamefont {J.}~\bibnamefont
  {Kang}}, \bibinfo {author} {\bibfnamefont {S.}~\bibnamefont {Tongay}},
  \bibinfo {author} {\bibfnamefont {J.}~\bibnamefont {Zhou}}, \bibinfo {author}
  {\bibfnamefont {J.}~\bibnamefont {Li}}, \ and\ \bibinfo {author}
  {\bibfnamefont {J.}~\bibnamefont {Wu}},\ }\href@noop {} {\bibfield  {journal}
  {\bibinfo  {journal} {Applied Physics Letters}\ }\textbf {\bibinfo {volume}
  {102}},\ \bibinfo {pages} {012111} (\bibinfo {year} {2013})}\BibitemShut
  {NoStop}%
\bibitem [{\citenamefont {Lang}\ and\ \citenamefont
  {Henry}(1978)}]{lang1978scanning}%
  \BibitemOpen
  \bibfield  {author} {\bibinfo {author} {\bibfnamefont {D.~V.}\ \bibnamefont
  {Lang}}\ and\ \bibinfo {author} {\bibfnamefont {C.~H.}\ \bibnamefont
  {Henry}},\ }\href@noop {} {\bibfield  {journal} {\bibinfo  {journal}
  {Solid-State Electronics}\ }\textbf {\bibinfo {volume} {21}},\ \bibinfo
  {pages} {1519} (\bibinfo {year} {1978})}\BibitemShut {NoStop}%
\bibitem [{\citenamefont {Ahn}\ \emph {et~al.}(2005)\citenamefont {Ahn},
  \citenamefont {Dunning},\ and\ \citenamefont {Park}}]{ahn2005scanning}%
  \BibitemOpen
  \bibfield  {author} {\bibinfo {author} {\bibfnamefont {Y.}~\bibnamefont
  {Ahn}}, \bibinfo {author} {\bibfnamefont {J.}~\bibnamefont {Dunning}}, \ and\
  \bibinfo {author} {\bibfnamefont {J.}~\bibnamefont {Park}},\ }\href@noop {}
  {\bibfield  {journal} {\bibinfo  {journal} {Nano Letters}\ }\textbf {\bibinfo
  {volume} {5}},\ \bibinfo {pages} {1367} (\bibinfo {year} {2005})}\BibitemShut
  {NoStop}%
\bibitem [{\citenamefont {Balasubramanian}\ \emph {et~al.}(2004)\citenamefont
  {Balasubramanian}, \citenamefont {Fan}, \citenamefont {Burghard},
  \citenamefont {Kern}, \citenamefont {Friedrich}, \citenamefont {Wannek},\
  and\ \citenamefont {Mews}}]{balasubramanian2004photoelectronic}%
  \BibitemOpen
  \bibfield  {author} {\bibinfo {author} {\bibfnamefont {K.}~\bibnamefont
  {Balasubramanian}}, \bibinfo {author} {\bibfnamefont {Y.}~\bibnamefont
  {Fan}}, \bibinfo {author} {\bibfnamefont {M.}~\bibnamefont {Burghard}},
  \bibinfo {author} {\bibfnamefont {K.}~\bibnamefont {Kern}}, \bibinfo {author}
  {\bibfnamefont {M.}~\bibnamefont {Friedrich}}, \bibinfo {author}
  {\bibfnamefont {U.}~\bibnamefont {Wannek}}, \ and\ \bibinfo {author}
  {\bibfnamefont {A.}~\bibnamefont {Mews}},\ }\href@noop {} {\bibfield
  {journal} {\bibinfo  {journal} {Applied Physics Letters}\ }\textbf {\bibinfo
  {volume} {84}},\ \bibinfo {pages} {2400} (\bibinfo {year}
  {2004})}\BibitemShut {NoStop}%
\bibitem [{\citenamefont {Gu}\ \emph {et~al.}(2006)\citenamefont {Gu},
  \citenamefont {Romankiewicz}, \citenamefont {David}, \citenamefont {Lensch},\
  and\ \citenamefont {Lauhon}}]{gu2006quantitative}%
  \BibitemOpen
  \bibfield  {author} {\bibinfo {author} {\bibfnamefont {Y.}~\bibnamefont
  {Gu}}, \bibinfo {author} {\bibfnamefont {J.~P.}\ \bibnamefont
  {Romankiewicz}}, \bibinfo {author} {\bibfnamefont {J.~K.}\ \bibnamefont
  {David}}, \bibinfo {author} {\bibfnamefont {J.~L.}\ \bibnamefont {Lensch}}, \
  and\ \bibinfo {author} {\bibfnamefont {L.~J.}\ \bibnamefont {Lauhon}},\
  }\href@noop {} {\bibfield  {journal} {\bibinfo  {journal} {Nano Letters}\
  }\textbf {\bibinfo {volume} {6}},\ \bibinfo {pages} {948} (\bibinfo {year}
  {2006})}\BibitemShut {NoStop}%
\bibitem [{\citenamefont {Yin}\ \emph {et~al.}(2012)\citenamefont {Yin},
  \citenamefont {Li}, \citenamefont {Li}, \citenamefont {Jiang}, \citenamefont
  {Shi}, \citenamefont {Sun}, \citenamefont {Lu}, \citenamefont {Zhang},
  \citenamefont {Chen},\ and\ \citenamefont {Zhang}}]{yin2011single}%
  \BibitemOpen
  \bibfield  {author} {\bibinfo {author} {\bibfnamefont {Z.}~\bibnamefont
  {Yin}}, \bibinfo {author} {\bibfnamefont {H.}~\bibnamefont {Li}}, \bibinfo
  {author} {\bibfnamefont {H.}~\bibnamefont {Li}}, \bibinfo {author}
  {\bibfnamefont {L.}~\bibnamefont {Jiang}}, \bibinfo {author} {\bibfnamefont
  {Y.}~\bibnamefont {Shi}}, \bibinfo {author} {\bibfnamefont {Y.}~\bibnamefont
  {Sun}}, \bibinfo {author} {\bibfnamefont {G.}~\bibnamefont {Lu}}, \bibinfo
  {author} {\bibfnamefont {Q.}~\bibnamefont {Zhang}}, \bibinfo {author}
  {\bibfnamefont {X.}~\bibnamefont {Chen}}, \ and\ \bibinfo {author}
  {\bibfnamefont {H.}~\bibnamefont {Zhang}},\ }\href@noop {} {\bibfield
  {journal} {\bibinfo  {journal} {ACS Nano}\ }\textbf {\bibinfo {volume} {6}},\
  \bibinfo {pages} {74} (\bibinfo {year} {2012})}\BibitemShut {NoStop}%
\bibitem [{\citenamefont {Perea-L{\'o}pez}\ \emph {et~al.}(2013)\citenamefont
  {Perea-L{\'o}pez}, \citenamefont {El{\'\i}as}, \citenamefont {Berkdemir},
  \citenamefont {Castro-Beltran}, \citenamefont {Guti{\'e}rrez}, \citenamefont
  {Feng}, \citenamefont {Lv}, \citenamefont {Hayashi}, \citenamefont
  {L{\'o}pez-Ur{\'\i}as}, \citenamefont {Ghosh}, \citenamefont {Muchharla},\
  and\ \citenamefont {Terrones}}]{perea2013photosensor}%
  \BibitemOpen
  \bibfield  {author} {\bibinfo {author} {\bibfnamefont {N.}~\bibnamefont
  {Perea-L{\'o}pez}}, \bibinfo {author} {\bibfnamefont {A.~L.}\ \bibnamefont
  {El{\'\i}as}}, \bibinfo {author} {\bibfnamefont {A.}~\bibnamefont
  {Berkdemir}}, \bibinfo {author} {\bibfnamefont {A.}~\bibnamefont
  {Castro-Beltran}}, \bibinfo {author} {\bibfnamefont {H.~R.}\ \bibnamefont
  {Guti{\'e}rrez}}, \bibinfo {author} {\bibfnamefont {S.}~\bibnamefont {Feng}},
  \bibinfo {author} {\bibfnamefont {R.}~\bibnamefont {Lv}}, \bibinfo {author}
  {\bibfnamefont {T.}~\bibnamefont {Hayashi}}, \bibinfo {author} {\bibfnamefont
  {F.}~\bibnamefont {L{\'o}pez-Ur{\'\i}as}}, \bibinfo {author} {\bibfnamefont
  {S.}~\bibnamefont {Ghosh}}, \bibinfo {author} {\bibfnamefont
  {B.}~\bibnamefont {Muchharla}}, \ and\ \bibinfo {author} {\bibfnamefont
  {M.}~\bibnamefont {Terrones}},\ }\href@noop {} {\bibfield  {journal}
  {\bibinfo  {journal} {Advanced Functional Materials}\ }\textbf {\bibinfo
  {volume} {23}},\ \bibinfo {pages} {5511} (\bibinfo {year}
  {2013})}\BibitemShut {NoStop}%
\bibitem [{\citenamefont {Choi}\ \emph {et~al.}(2012)\citenamefont {Choi},
  \citenamefont {Cho}, \citenamefont {Konar}, \citenamefont {Lee},
  \citenamefont {Cha}, \citenamefont {Hong}, \citenamefont {Kim}, \citenamefont
  {Kim}, \citenamefont {Jena}, \citenamefont {Joo},\ and\ \citenamefont
  {Kim}}]{choi2012high}%
  \BibitemOpen
  \bibfield  {author} {\bibinfo {author} {\bibfnamefont {W.}~\bibnamefont
  {Choi}}, \bibinfo {author} {\bibfnamefont {M.~Y.}\ \bibnamefont {Cho}},
  \bibinfo {author} {\bibfnamefont {A.}~\bibnamefont {Konar}}, \bibinfo
  {author} {\bibfnamefont {J.~H.}\ \bibnamefont {Lee}}, \bibinfo {author}
  {\bibfnamefont {G.-B.}\ \bibnamefont {Cha}}, \bibinfo {author} {\bibfnamefont
  {S.~C.}\ \bibnamefont {Hong}}, \bibinfo {author} {\bibfnamefont
  {S.}~\bibnamefont {Kim}}, \bibinfo {author} {\bibfnamefont {J.}~\bibnamefont
  {Kim}}, \bibinfo {author} {\bibfnamefont {D.}~\bibnamefont {Jena}}, \bibinfo
  {author} {\bibfnamefont {J.}~\bibnamefont {Joo}}, \ and\ \bibinfo {author}
  {\bibfnamefont {S.}~\bibnamefont {Kim}},\ }\href@noop {} {\bibfield
  {journal} {\bibinfo  {journal} {Advanced Materials}\ }\textbf {\bibinfo
  {volume} {24}},\ \bibinfo {pages} {5832} (\bibinfo {year}
  {2012})}\BibitemShut {NoStop}%
\bibitem [{\citenamefont {Zhao}\ \emph {et~al.}(2013)\citenamefont {Zhao},
  \citenamefont {Ghorannevis}, \citenamefont {Chu}, \citenamefont {Toh},
  \citenamefont {Kloc}, \citenamefont {Tan},\ and\ \citenamefont
  {Eda}}]{zhao_evolution_2013}%
  \BibitemOpen
  \bibfield  {author} {\bibinfo {author} {\bibfnamefont {W.}~\bibnamefont
  {Zhao}}, \bibinfo {author} {\bibfnamefont {Z.}~\bibnamefont {Ghorannevis}},
  \bibinfo {author} {\bibfnamefont {L.}~\bibnamefont {Chu}}, \bibinfo {author}
  {\bibfnamefont {M.}~\bibnamefont {Toh}}, \bibinfo {author} {\bibfnamefont
  {C.}~\bibnamefont {Kloc}}, \bibinfo {author} {\bibfnamefont {P.-H.}\
  \bibnamefont {Tan}}, \ and\ \bibinfo {author} {\bibfnamefont
  {G.}~\bibnamefont {Eda}},\ }\href {\doibase 10.1021/nn305275h} {\bibfield
  {journal} {\bibinfo  {journal} {{ACS} Nano}\ }\textbf {\bibinfo {volume}
  {7}},\ \bibinfo {pages} {791} (\bibinfo {year} {2013})}\BibitemShut {NoStop}%
\bibitem [{\citenamefont {Jiang}(2012)}]{jiang_electronic_2012}%
  \BibitemOpen
  \bibfield  {author} {\bibinfo {author} {\bibfnamefont {H.}~\bibnamefont
  {Jiang}},\ }\href {\doibase 10.1021/jp300079d} {\bibfield  {journal}
  {\bibinfo  {journal} {The Journal of Physical Chemistry C}\ }\textbf
  {\bibinfo {volume} {116}},\ \bibinfo {pages} {7664} (\bibinfo {year}
  {2012})}\BibitemShut {NoStop}%
\bibitem [{\citenamefont {Ballif}\ \emph {et~al.}(1996)\citenamefont {Ballif},
  \citenamefont {Regula}, \citenamefont {Schmid}, \citenamefont
  {Rem{\v{s}}kar}, \citenamefont {Sanjines},\ and\ \citenamefont
  {L{\'e}vy}}]{ballif1996preparation}%
  \BibitemOpen
  \bibfield  {author} {\bibinfo {author} {\bibfnamefont {C.}~\bibnamefont
  {Ballif}}, \bibinfo {author} {\bibfnamefont {M.}~\bibnamefont {Regula}},
  \bibinfo {author} {\bibfnamefont {P.}~\bibnamefont {Schmid}}, \bibinfo
  {author} {\bibfnamefont {M.}~\bibnamefont {Rem{\v{s}}kar}}, \bibinfo {author}
  {\bibfnamefont {R.}~\bibnamefont {Sanjines}}, \ and\ \bibinfo {author}
  {\bibfnamefont {F.}~\bibnamefont {L{\'e}vy}},\ }\href@noop {} {\bibfield
  {journal} {\bibinfo  {journal} {Applied Physics A}\ }\textbf {\bibinfo
  {volume} {62}},\ \bibinfo {pages} {543} (\bibinfo {year} {1996})}\BibitemShut
  {NoStop}%
\bibitem [{\citenamefont {Fu}\ \emph {et~al.}(2011)\citenamefont {Fu},
  \citenamefont {Zou}, \citenamefont {Wang}, \citenamefont {Zhang},
  \citenamefont {Yu},\ and\ \citenamefont {Wu}}]{fu_electrothermal_2011}%
  \BibitemOpen
  \bibfield  {author} {\bibinfo {author} {\bibfnamefont {D.}~\bibnamefont
  {Fu}}, \bibinfo {author} {\bibfnamefont {J.}~\bibnamefont {Zou}}, \bibinfo
  {author} {\bibfnamefont {K.}~\bibnamefont {Wang}}, \bibinfo {author}
  {\bibfnamefont {R.}~\bibnamefont {Zhang}}, \bibinfo {author} {\bibfnamefont
  {D.}~\bibnamefont {Yu}}, \ and\ \bibinfo {author} {\bibfnamefont
  {J.}~\bibnamefont {Wu}},\ }\href {\doibase 10.1021/nl2018806} {\bibfield
  {journal} {\bibinfo  {journal} {Nano Letters}\ }\textbf {\bibinfo {volume}
  {11}},\ \bibinfo {pages} {3809} (\bibinfo {year} {2011})}\BibitemShut
  {NoStop}%
\bibitem [{\citenamefont {Kim}\ \emph {et~al.}(2010)\citenamefont {Kim},
  \citenamefont {Lee}, \citenamefont {Cho}, \citenamefont {Kang},\ and\
  \citenamefont {Jo}}]{kim2010diameter}%
  \BibitemOpen
  \bibfield  {author} {\bibinfo {author} {\bibfnamefont {C.-J.}\ \bibnamefont
  {Kim}}, \bibinfo {author} {\bibfnamefont {H.-S.}\ \bibnamefont {Lee}},
  \bibinfo {author} {\bibfnamefont {Y.-J.}\ \bibnamefont {Cho}}, \bibinfo
  {author} {\bibfnamefont {K.}~\bibnamefont {Kang}}, \ and\ \bibinfo {author}
  {\bibfnamefont {M.-H.}\ \bibnamefont {Jo}},\ }\href@noop {} {\bibfield
  {journal} {\bibinfo  {journal} {Nano Letters}\ }\textbf {\bibinfo {volume}
  {10}},\ \bibinfo {pages} {2043} (\bibinfo {year} {2010})}\BibitemShut
  {NoStop}%
\bibitem [{\citenamefont {Ubrig}\ \emph {et~al.}(2014)\citenamefont {Ubrig},
  \citenamefont {Jo}, \citenamefont {Berger}, \citenamefont {Morpurgo},\ and\
  \citenamefont {Kuzmenko}}]{ubrig2014scanning}%
  \BibitemOpen
  \bibfield  {author} {\bibinfo {author} {\bibfnamefont {N.}~\bibnamefont
  {Ubrig}}, \bibinfo {author} {\bibfnamefont {S.}~\bibnamefont {Jo}}, \bibinfo
  {author} {\bibfnamefont {H.}~\bibnamefont {Berger}}, \bibinfo {author}
  {\bibfnamefont {A.~F.}\ \bibnamefont {Morpurgo}}, \ and\ \bibinfo {author}
  {\bibfnamefont {A.~B.}\ \bibnamefont {Kuzmenko}},\ }\href@noop {} {\bibfield
  {journal} {\bibinfo  {journal} {Applied Physics Letters}\ }\textbf {\bibinfo
  {volume} {104}},\ \bibinfo {pages} {171112} (\bibinfo {year}
  {2014})}\BibitemShut {NoStop}%
\bibitem [{\citenamefont {Braga}\ \emph {et~al.}(2012)\citenamefont {Braga},
  \citenamefont {Gutie�rrez~Lezama}, \citenamefont {Berger},\ and\
  \citenamefont {Morpurgo}}]{braga2012quantitative}%
  \BibitemOpen
  \bibfield  {author} {\bibinfo {author} {\bibfnamefont {D.}~\bibnamefont
  {Braga}}, \bibinfo {author} {\bibfnamefont {I.}~\bibnamefont
  {Gutie�rrez~Lezama}}, \bibinfo {author} {\bibfnamefont {H.}~\bibnamefont
  {Berger}}, \ and\ \bibinfo {author} {\bibfnamefont {A.~F.}\ \bibnamefont
  {Morpurgo}},\ }\href@noop {} {\bibfield  {journal} {\bibinfo  {journal} {Nano
  Letters}\ }\textbf {\bibinfo {volume} {12}},\ \bibinfo {pages} {5218}
  (\bibinfo {year} {2012})}\BibitemShut {NoStop}%
\bibitem [{\citenamefont {Shi}\ \emph {et~al.}(2013)\citenamefont {Shi},
  \citenamefont {Yan}, \citenamefont {Bertolazzi}, \citenamefont {Brivio},
  \citenamefont {Gao}, \citenamefont {Kis}, \citenamefont {Jena}, \citenamefont
  {Xing},\ and\ \citenamefont {Huang}}]{shi2013exciton}%
  \BibitemOpen
  \bibfield  {author} {\bibinfo {author} {\bibfnamefont {H.}~\bibnamefont
  {Shi}}, \bibinfo {author} {\bibfnamefont {R.}~\bibnamefont {Yan}}, \bibinfo
  {author} {\bibfnamefont {S.}~\bibnamefont {Bertolazzi}}, \bibinfo {author}
  {\bibfnamefont {J.}~\bibnamefont {Brivio}}, \bibinfo {author} {\bibfnamefont
  {B.}~\bibnamefont {Gao}}, \bibinfo {author} {\bibfnamefont {A.}~\bibnamefont
  {Kis}}, \bibinfo {author} {\bibfnamefont {D.}~\bibnamefont {Jena}}, \bibinfo
  {author} {\bibfnamefont {H.~G.}\ \bibnamefont {Xing}}, \ and\ \bibinfo
  {author} {\bibfnamefont {L.}~\bibnamefont {Huang}},\ }\href@noop {}
  {\bibfield  {journal} {\bibinfo  {journal} {ACS Nano}\ }\textbf {\bibinfo
  {volume} {7}},\ \bibinfo {pages} {1072} (\bibinfo {year} {2013})}\BibitemShut
  {NoStop}%
\bibitem [{\citenamefont {Mouri}\ \emph {et~al.}(2013)\citenamefont {Mouri},
  \citenamefont {Miyauchi},\ and\ \citenamefont {Matsuda}}]{mouri2013tunable}%
  \BibitemOpen
  \bibfield  {author} {\bibinfo {author} {\bibfnamefont {S.}~\bibnamefont
  {Mouri}}, \bibinfo {author} {\bibfnamefont {Y.}~\bibnamefont {Miyauchi}}, \
  and\ \bibinfo {author} {\bibfnamefont {K.}~\bibnamefont {Matsuda}},\
  }\href@noop {} {\bibfield  {journal} {\bibinfo  {journal} {Nano Letters}\
  }\textbf {\bibinfo {volume} {13}},\ \bibinfo {pages} {5944} (\bibinfo {year}
  {2013})}\BibitemShut {NoStop}%
\bibitem [{\citenamefont {Freitag}\ \emph
  {et~al.}(2007{\natexlab{a}})\citenamefont {Freitag}, \citenamefont {Tsang},
  \citenamefont {Bol}, \citenamefont {Yuan}, \citenamefont {Liu},\ and\
  \citenamefont {Avouris}}]{freitag2007imaging}%
  \BibitemOpen
  \bibfield  {author} {\bibinfo {author} {\bibfnamefont {M.}~\bibnamefont
  {Freitag}}, \bibinfo {author} {\bibfnamefont {J.~C.}\ \bibnamefont {Tsang}},
  \bibinfo {author} {\bibfnamefont {A.}~\bibnamefont {Bol}}, \bibinfo {author}
  {\bibfnamefont {D.}~\bibnamefont {Yuan}}, \bibinfo {author} {\bibfnamefont
  {J.}~\bibnamefont {Liu}}, \ and\ \bibinfo {author} {\bibfnamefont
  {P.}~\bibnamefont {Avouris}},\ }\href@noop {} {\bibfield  {journal} {\bibinfo
   {journal} {Nano Letters}\ }\textbf {\bibinfo {volume} {7}},\ \bibinfo
  {pages} {2037} (\bibinfo {year} {2007}{\natexlab{a}})}\BibitemShut {NoStop}%
\bibitem [{\citenamefont {Lee}\ \emph {et~al.}(2007)\citenamefont {Lee},
  \citenamefont {Balasubramanian}, \citenamefont {Dorfm{\"u}ller},
  \citenamefont {Vogelgesang}, \citenamefont {Fu}, \citenamefont {Mews},
  \citenamefont {Burghard},\ and\ \citenamefont {Kern}}]{lee2007electronic}%
  \BibitemOpen
  \bibfield  {author} {\bibinfo {author} {\bibfnamefont {E.~J.}\ \bibnamefont
  {Lee}}, \bibinfo {author} {\bibfnamefont {K.}~\bibnamefont
  {Balasubramanian}}, \bibinfo {author} {\bibfnamefont {J.}~\bibnamefont
  {Dorfm{\"u}ller}}, \bibinfo {author} {\bibfnamefont {R.}~\bibnamefont
  {Vogelgesang}}, \bibinfo {author} {\bibfnamefont {N.}~\bibnamefont {Fu}},
  \bibinfo {author} {\bibfnamefont {A.}~\bibnamefont {Mews}}, \bibinfo {author}
  {\bibfnamefont {M.}~\bibnamefont {Burghard}}, \ and\ \bibinfo {author}
  {\bibfnamefont {K.}~\bibnamefont {Kern}},\ }\href@noop {} {\bibfield
  {journal} {\bibinfo  {journal} {Small}\ }\textbf {\bibinfo {volume} {3}},\
  \bibinfo {pages} {2038} (\bibinfo {year} {2007})}\BibitemShut {NoStop}%
\bibitem [{\citenamefont {Zhang}\ \emph
  {et~al.}(2012{\natexlab{b}})\citenamefont {Zhang}, \citenamefont {Ning},
  \citenamefont {Liu}, \citenamefont {Xu}, \citenamefont {Guo}, \citenamefont
  {Zak}, \citenamefont {Zhang}, \citenamefont {Wang}, \citenamefont {Tenne},\
  and\ \citenamefont {Chen}}]{zhang2012electrical}%
  \BibitemOpen
  \bibfield  {author} {\bibinfo {author} {\bibfnamefont {C.}~\bibnamefont
  {Zhang}}, \bibinfo {author} {\bibfnamefont {Z.}~\bibnamefont {Ning}},
  \bibinfo {author} {\bibfnamefont {Y.}~\bibnamefont {Liu}}, \bibinfo {author}
  {\bibfnamefont {T.}~\bibnamefont {Xu}}, \bibinfo {author} {\bibfnamefont
  {Y.}~\bibnamefont {Guo}}, \bibinfo {author} {\bibfnamefont {A.}~\bibnamefont
  {Zak}}, \bibinfo {author} {\bibfnamefont {Z.}~\bibnamefont {Zhang}}, \bibinfo
  {author} {\bibfnamefont {S.}~\bibnamefont {Wang}}, \bibinfo {author}
  {\bibfnamefont {R.}~\bibnamefont {Tenne}}, \ and\ \bibinfo {author}
  {\bibfnamefont {Q.}~\bibnamefont {Chen}},\ }\href@noop {} {\bibfield
  {journal} {\bibinfo  {journal} {Applied Physics Letters}\ }\textbf {\bibinfo
  {volume} {101}},\ \bibinfo {pages} {113112} (\bibinfo {year}
  {2012}{\natexlab{b}})}\BibitemShut {NoStop}%
\bibitem [{\citenamefont {Freitag}\ \emph
  {et~al.}(2007{\natexlab{b}})\citenamefont {Freitag}, \citenamefont {Tsang},
  \citenamefont {Bol}, \citenamefont {Avouris}, \citenamefont {Yuan},\ and\
  \citenamefont {Liu}}]{freitag2007scanning}%
  \BibitemOpen
  \bibfield  {author} {\bibinfo {author} {\bibfnamefont {M.}~\bibnamefont
  {Freitag}}, \bibinfo {author} {\bibfnamefont {J.~C.}\ \bibnamefont {Tsang}},
  \bibinfo {author} {\bibfnamefont {A.}~\bibnamefont {Bol}}, \bibinfo {author}
  {\bibfnamefont {P.}~\bibnamefont {Avouris}}, \bibinfo {author} {\bibfnamefont
  {D.}~\bibnamefont {Yuan}}, \ and\ \bibinfo {author} {\bibfnamefont
  {J.}~\bibnamefont {Liu}},\ }\href@noop {} {\bibfield  {journal} {\bibinfo
  {journal} {Applied Physics Letters}\ }\textbf {\bibinfo {volume} {91}},\
  \bibinfo {pages} {031101} (\bibinfo {year} {2007}{\natexlab{b}})}\BibitemShut
  {NoStop}%
\bibitem [{\citenamefont {Balasubramanian}\ \emph {et~al.}(2005)\citenamefont
  {Balasubramanian}, \citenamefont {Burghard}, \citenamefont {Kern},
  \citenamefont {Scolari},\ and\ \citenamefont
  {Mews}}]{balasubramanian2005photocurrent}%
  \BibitemOpen
  \bibfield  {author} {\bibinfo {author} {\bibfnamefont {K.}~\bibnamefont
  {Balasubramanian}}, \bibinfo {author} {\bibfnamefont {M.}~\bibnamefont
  {Burghard}}, \bibinfo {author} {\bibfnamefont {K.}~\bibnamefont {Kern}},
  \bibinfo {author} {\bibfnamefont {M.}~\bibnamefont {Scolari}}, \ and\
  \bibinfo {author} {\bibfnamefont {A.}~\bibnamefont {Mews}},\ }\href@noop {}
  {\bibfield  {journal} {\bibinfo  {journal} {Nano Letters}\ }\textbf {\bibinfo
  {volume} {5}},\ \bibinfo {pages} {507} (\bibinfo {year} {2005})}\BibitemShut
  {NoStop}%
\bibitem [{\citenamefont {Yu}\ \emph {et~al.}(2009)\citenamefont {Yu},
  \citenamefont {Zhao}, \citenamefont {Ryu}, \citenamefont {Brus},
  \citenamefont {Kim},\ and\ \citenamefont {Kim}}]{yu_tuning_2009}%
  \BibitemOpen
  \bibfield  {author} {\bibinfo {author} {\bibfnamefont {Y.-J.}\ \bibnamefont
  {Yu}}, \bibinfo {author} {\bibfnamefont {Y.}~\bibnamefont {Zhao}}, \bibinfo
  {author} {\bibfnamefont {S.}~\bibnamefont {Ryu}}, \bibinfo {author}
  {\bibfnamefont {L.~E.}\ \bibnamefont {Brus}}, \bibinfo {author}
  {\bibfnamefont {K.~S.}\ \bibnamefont {Kim}}, \ and\ \bibinfo {author}
  {\bibfnamefont {P.}~\bibnamefont {Kim}},\ }\href {\doibase 10.1021/nl901572a}
  {\bibfield  {journal} {\bibinfo  {journal} {Nano Letters}\ }\textbf {\bibinfo
  {volume} {9}},\ \bibinfo {pages} {3430} (\bibinfo {year} {2009})}\BibitemShut
  {NoStop}%
\bibitem [{\citenamefont {Nan}\ \emph {et~al.}(2014)\citenamefont {Nan},
  \citenamefont {Wang}, \citenamefont {Wang}, \citenamefont {Liang},
  \citenamefont {Lu}, \citenamefont {Chen}, \citenamefont {He}, \citenamefont
  {Tan}, \citenamefont {Miao}, \citenamefont {Wang}, \citenamefont {Wang},\
  and\ \citenamefont {Ni}}]{nan2014strong}%
  \BibitemOpen
  \bibfield  {author} {\bibinfo {author} {\bibfnamefont {H.}~\bibnamefont
  {Nan}}, \bibinfo {author} {\bibfnamefont {Z.}~\bibnamefont {Wang}}, \bibinfo
  {author} {\bibfnamefont {W.}~\bibnamefont {Wang}}, \bibinfo {author}
  {\bibfnamefont {Z.}~\bibnamefont {Liang}}, \bibinfo {author} {\bibfnamefont
  {Y.}~\bibnamefont {Lu}}, \bibinfo {author} {\bibfnamefont {Q.}~\bibnamefont
  {Chen}}, \bibinfo {author} {\bibfnamefont {D.}~\bibnamefont {He}}, \bibinfo
  {author} {\bibfnamefont {P.}~\bibnamefont {Tan}}, \bibinfo {author}
  {\bibfnamefont {F.}~\bibnamefont {Miao}}, \bibinfo {author} {\bibfnamefont
  {X.}~\bibnamefont {Wang}}, \bibinfo {author} {\bibfnamefont {J.}~\bibnamefont
  {Wang}}, \ and\ \bibinfo {author} {\bibfnamefont {Z.}~\bibnamefont {Ni}},\
  }\href {\doibase 10.1021/nn500532f} {\bibfield  {journal} {\bibinfo
  {journal} {ACS Nano}\ }\textbf {\bibinfo {volume} {8}},\ \bibinfo {pages}
  {5738} (\bibinfo {year} {2014})}\BibitemShut {NoStop}%
\bibitem [{\citenamefont {Bockrath}\ \emph {et~al.}(2001)\citenamefont
  {Bockrath}, \citenamefont {Liang}, \citenamefont {Bozovic}, \citenamefont
  {Hafner}, \citenamefont {Lieber}, \citenamefont {Tinkham},\ and\
  \citenamefont {Park}}]{bockrath_resonant_2001}%
  \BibitemOpen
  \bibfield  {author} {\bibinfo {author} {\bibfnamefont {M.}~\bibnamefont
  {Bockrath}}, \bibinfo {author} {\bibfnamefont {W.}~\bibnamefont {Liang}},
  \bibinfo {author} {\bibfnamefont {D.}~\bibnamefont {Bozovic}}, \bibinfo
  {author} {\bibfnamefont {J.~H.}\ \bibnamefont {Hafner}}, \bibinfo {author}
  {\bibfnamefont {C.~M.}\ \bibnamefont {Lieber}}, \bibinfo {author}
  {\bibfnamefont {M.}~\bibnamefont {Tinkham}}, \ and\ \bibinfo {author}
  {\bibfnamefont {H.}~\bibnamefont {Park}},\ }\href {\doibase
  10.1126/science.291.5502.283} {\bibfield  {journal} {\bibinfo  {journal}
  {Science}\ }\textbf {\bibinfo {volume} {291}},\ \bibinfo {pages} {283}
  (\bibinfo {year} {2001})}\BibitemShut {NoStop}%
\end{thebibliography}
\end{document}